\documentclass[11pt]{article}

\include{graphicx}
\begin{document}

\begin{center}
\textbf{A new model for competition between many languages}

\c{C}a\u{g}lar Tuncay

Department of Physics, Middle East Technical University

06531 Ankara, Turkey

caglart@metu.edu.tr
\end{center}

\textbf{Abstract:} Time evolutions of number of cities, population of
cities, world population, and size distribution of present languages are
studied in terms of a new model, where population of each city increases by
a random rate and decreases by a random division. World population and size
distribution of languages come out in good agreement with the available 
empirical data.

\textbf{Introduction:} Many old languages, like ancient Greek and Roman may
have spread in terms of colonization, where the cities were shaped by the
current environmental conditions (see Viviane model, Refs.1, 2, and
references therein). Yet, the cited mechanism may not cover all the
relevant situations. Many medieval languages of the empires are not spoken
on the same lands presently. On the other hand, Chinese and Indian languages
might have spread in terms of increasing population, rather than
colonization and other means. In the present contribution a new model for
many languages is suggested, where city population increases by a random rate
and decreases by a random division.[3,4] The model is given in the
following section; applications and results are displayed in next one. The last
section is devoted for discussion and conclusion.

\textbf{Model:} We start with some number of cities ($M(t)$), which
varies in time ($t$). Each city ($i$) might have equal initial population 
($P_i(t=0)$) or the initial population of each may be random and equal to
$Pr_i$, where $P$ is some constant and $r_i$ is a random real number $0 \leq r_i< 1$, with $i\leq M(0)$.
We assign a random population growth rate ($R_i$) to each
society, $R_i=Rr_i$, where $R$ is constant. Through the
evolution of history, each city at each time step with probability $H$ gives
birth to a new one where the splitting ratio of population equals $S$, 
which is assumed to be the same for all the cities. If the current
population of the city ($i$) is $P_i(t)$, $SP_i(t)$ many citizens move
away to establish a new city and $(1-S)P_i(t)$ many remain. And, due to
the present splitting, the number of cities $M(t)$ increases by one; if two
cities split at $t$, then $M(t)$ increases by two, etc. Please note that,
results do not change if $1-S$ is substituted for $S$, i.e., if the moved
and remained citizens are interchanged. So, the greatest value for $S$ is
0.5, effectively.

Whenever a new city is established, her people may survive the previous
language or create a new one. We assume that new conditions (geographical,
etc) and lack of interactions with the home city promotes a new language.
So, we take one language for each city.

Furthermore, cities may be considered as countries (states) equally well,
since almost each city was a state in past. And for $t \to \infty$, these cities
which have big population may obviously be taken as a
state (country). In any case, the unit land ($i$), i.e., city (country,
state), may be evaluated as the totality of humans speaking the same language
($i$), and our assumption of one language for one city is satisfied. So, $i$
counts cities and languages.

\underline{\textit{Initiation}}$:$ We assume $M(0)$ many cities existed
initially, and each may be assumed to have more or less the same population,
at least in order of magnitude. Yet, to study the effect of small cities, we
consider random initial populations too, and assign $Pr_i$ many citizens
to each, where $P$ is some constant and $r_i$ is a random real number $0
\leq r_i<1$, with $i\leq M(0)$. The opulation growth rate ($R_i$) is
also fixed initially, and not varied in time. (Further generation cities
randomly get a new $R_i$ during splitting, and do not change this parameter
later.)

\underline {\textit{Evolution}}$:$ We let the cities grow in time, within a
process known as multiplicative noise,

$$P_i(t)=(1+R_i)P_i(t-1) \quad , \eqno(1) $$

\noindent and if a random number is smaller than the splitting probability $
H $ the city ($i$) splits.

The world population ($W$) is,

$$W(t)=\sum_{i=1}^{M(t)}P_i(t) \quad , \eqno(2)$$

\noindent and the model must predict the real data for $W(t)$ [5,6].

Please note that, the introduced parameters have units involving time, and our
time unit is arbitrary here. And, after some period of evolution in time we
(reaching the present) stop the simulation and calculate the probability
distribution function (PDF) for the number of cities and size.

\textbf{Applications and Results:} The basic parameters are: $M(0)$ (initial
number of cities), $P_i(t=0)$ (=$Pr_i$, initial population of each city,
and uniqueness ($r_i=1$, for all $i$) or randomness of it ($0 \leq r_i < 1$, 
for all $i$)), $R_i \;(=Rr_i$, population growth rate), and $H$ (historical
factor for splitting of cities, assumed to be the same for all the
cities). The splitting ratio $S$ is also considered as universal. Some of the
pronounced parameters would be eliminated if we knew the real historical
data. Please note that, $M(0)$ and $P(0)$ define the origin of our time
scale, and an increase in $M(0)$ and in $P(0)$ means shifting the time origin
forward, and vice versa. On the other hand, $R$ and $H$ are defined per unit
of time. So scaling of only one of them means scaling the time axis by the
same factor, but inversely, where results remain invariant.

We define our time unit to be one decade (10 years), we take $t=0$ at 10,000
B.C. and run simulations for 12,000 years, i.e., for 1,200 points.

\underline{\textit{Pre-historic world: }}We don't know world population ($%
W(0)$) and number of cities ($M(0)$) at 10,000 B.C., and $R_{i}$ is also
unknown. We run our simulations for various $M(0)$, $P_{i}(0)$, and $R_{i}$,
and tried to predict the real data for $W(t)$ (Figure 1, where earlier
portion is obtained by estimation [5]). In Fig. 1 the super-exponential 
behavior in $W$ is crucial. We consider also the prediction
(made by United Nations) about world population to be 6.5 billion in 2005,
and to be about 10 billion in 2050 [6]. We display one of our
results for $W(t)$ in Figure 2, where super-exponential character may be
observed, as explained within the caption.

In figures 2,3 we take $W(0)=500,000$ by guess. Accordingly, we take $M(0)=1000$ with
$P=10,000$ and we define $P_i(0)$ randomly. Furthermore, for splitting $S$ may
be taken about 10 \% . Yet, when a country splits, $S$ may vary from about 50 \%
to 10 \%. Please note that, $H$ may be considered as the rate of increasing 
the number of cities; variation of $H$ changes $M(t)$ for a given $W(t)$.

Within the present approach, we get an exponential growth in $M(t)$, and
super-exponential growth in $W(t)$. $H$ and $P$ are effective on the rate of 
$M(t)$ and $W(t)$, respectively. So, for a given initial world, one may
have a variety in $M(t)$ for a given $W(t)$, and vice versa.

In Figure 4 is displayed the PDF for the current number of cities 
or languages where we assumed initially one million people, 
now all speaking the same language. For comparison, 
Figure 5 is the empirical PDF for the current
number of cities or languages [7]. 

Within our results, (the population of the biggest city or) the number of people
speaking the most wide-spread language and the current world population came out
as $1.025 \times 10^9$, and $13.3 \times 10^9$ respectively, so the ratio is 
7.7 \%. 

In all the figures we 
utilized $S=0.5$, which means that the cities are divided by a half. In order
to see the effect of small $S$, i.e. fragmentation of cities, we try $S=0.1$, 
0.01, 0.001, and 0.0001 (not shown).

\textbf{Discussion and Conclusion:} Many parameters of the present formalism
are not crucial for the size distribution of languages in Fig. 4; and, one
of the parameters could be dropped, i.e. absorbed with the unit for time. In
our runs, several sets of parameters (with minor changes) gave similar
results to Figure 4, and we selected one of them for display here. What
is most important for the size distribution of languages is the final
population of each city. The present model and the method may be considered as a
reasonable one, since the reality may be predicted. Furthermore, we do not
have any contradiction with Viviane model and Schulze model [8].

\textbf{References}
\parindent 0pt

[1] Schulze, C. and Stauffer, D., Computer simulation of language
competition by physicists. In: Chakrabarti, B. K., A. Chakraborti and A.
Chatterjee (eds.), Econophysics and Sociophysics: Trends and Perspectives.
Weinheim: WILEY-VCH Verlag (2006); and: Recent developments in computer
simulations of language competition, Computing in Science and Engineering 8
(May/June) 86-93.

[2] S. Wichmann, D. Stauffer, F.W.S. Lima, and C. Schulze C., Modelling
linguistic taxonomic dynamics, accepted by Trans. Philological Soc. (2007);
arXiv:physics/0604146.

[3] H.A. Simon, Models of Man, Wiley, New York 1957

[4] V. Novotny and P. Drozdz, Proc. Roy. Soc. London \textbf{B267}, 947, (2000).

[5] A. Johansen , D. Sornette, Physica A \textbf{294}, 465 (2001).

[6] URL: http://www.un.org/esa/population/publications/WPP2004/

2004Highlights{\_}finalrevised.pdf

[7] Based on: B.F. Grimes, {\it Ethnologue: Languages of the World}
(14th edn. 2000). Dallas, TX: Summer Institute of Linguistics.

[8] C. Schulze, D. Stauffer, Int. J. Mod. Phys. C \textbf{16}, 781
(2005).

\begin{figure}[hbt]
\begin{center}
\includegraphics[scale=0.8]{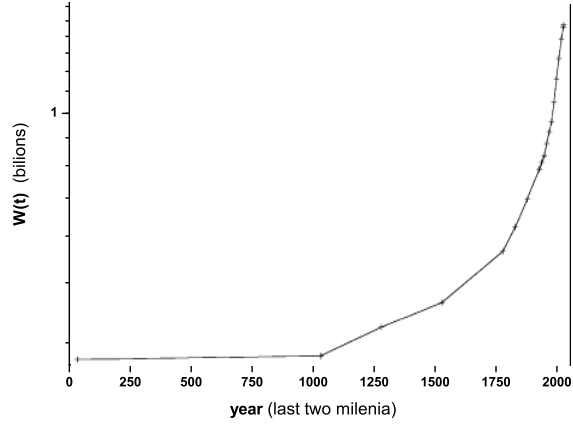}
\end{center}
\caption{
Real empirical data for $W(t)$, where earlier portion is
obtained by estimation (see also [5]).
}
\end{figure}
\begin{figure}[hbt]
\begin{center}
\includegraphics[scale=0.8]{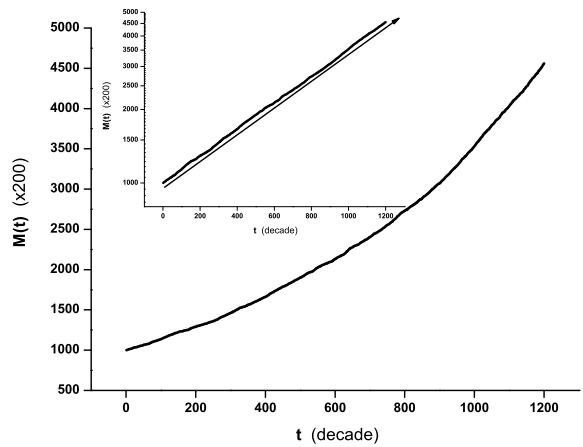}
\end{center}
\caption{
Evolution of $M(t)$, where the parameters are: $M(0)$%
=1000, $R=0.0024$, $P=10,000$, $R_i(0)=0.01 r_i$ ($0\leq r_i<1$), 
$S_i=0.5$, $H=0.0013$. Please note the exponential growth as the arrow
indicates within the inset.
}
\end{figure}
\begin{figure}[hbt]
\begin{center}
\includegraphics[scale=0.8]{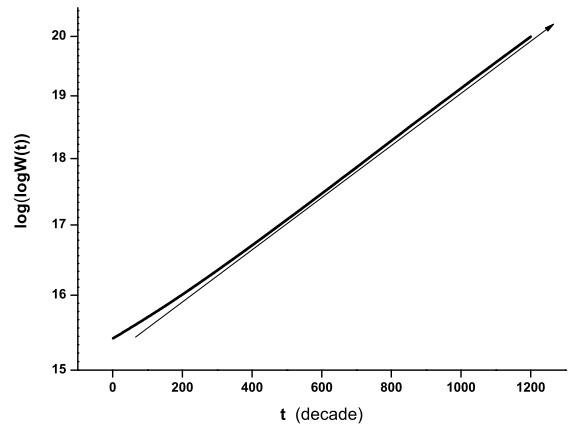}
\end{center}
\caption{
Evolution of $\log(\log(W(t)))$, with the same parameters as
in Fig. 3. Please note the super-exponential growing in terms of the slope,
as the arrow designates.
}
\end{figure}
\begin{figure}[hbt]
\begin{center}
\includegraphics[scale=0.8]{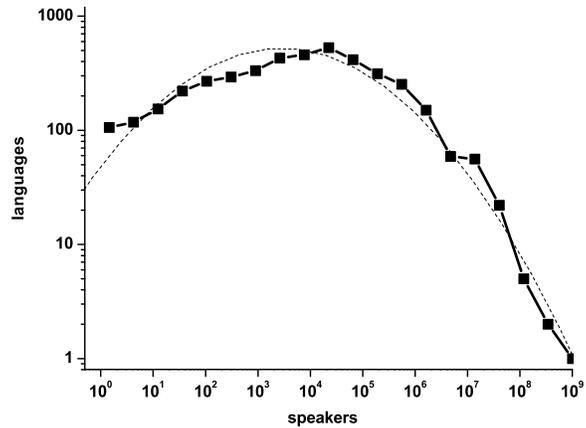}
\end{center}
\caption{
PDF for the number of languages at present. Here, we have
initially one million people speaking the same language. At present (in 1300
tours) the size of the biggest language is $1.025\times 10^9$ and the world
population is $13.3\times 10^9$; so, the ratio is 0.07 (12.9:1). Other
parameters are: $H=0.0021,\; S=0.5,\; R=0.02$. The dashed curve
demonstrates that, our distribution is a slightly asymmetric Gaussian. (See
also the relevant text.)
}
\end{figure}
\begin{figure}[hbt]
\begin{center}
\includegraphics[scale=0.8]{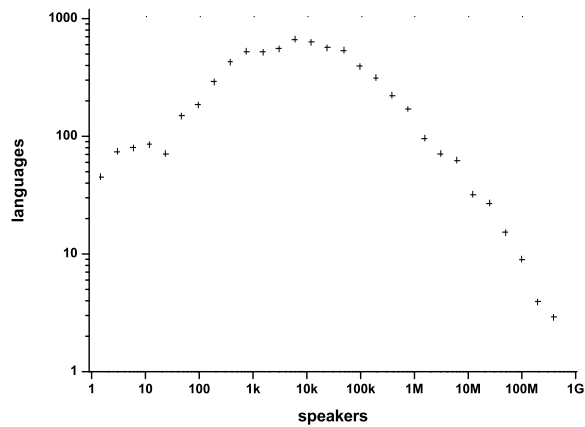}
\end{center}
\caption{
Empirical PDF for the current number cities or languages [8].
}
\end{figure}

\end{document}